\documentclass[12pt]{article}
\usepackage{amsmath}
\begin{document}
\begin{flushright}
KANAZAWA-13-01\\
April, 2013
\end{flushright}
\vspace*{1cm}

\begin{center} 
{\Large\bf Leptogenesis and dark matter detection in a TeV scale 
neutrino mass model with inverted mass hierarchy}
\vspace*{1cm}

{\Large Shoichi Kashiwase}\footnote{e-mail:~shoichi@hep.s.
kanazawa-u.ac.jp}
{\Large ~and ~Daijiro Suematsu}\footnote{e-mail:~suematsu@hep.
s.kanazawa-u.ac.jp}
\vspace*{1cm}\\

{\it Institute for Theoretical Physics, Kanazawa University, 
\\ Kanazawa 920-1192, Japan}
\end{center}
\vspace*{1.5cm} 

\noindent
{\Large\bf Abstract}\\
Realization of the inverted hierarchy is studied in the radiative
neutrino mass model with an additional doublet, in which neutrino masses 
and dark matter could be induced from a common particle. 
We show that the sufficient baryon number asymmetry is generated through 
resonant leptogenesis even for the case with rather mild degeneracy
among TeV scale right-handed neutrinos. We also discuss the relation
between this neutrino mass generation mechanism and low energy
experiments for the DM direct search, the neutrinoless double $\beta$ 
decay and so on. 
      
\newpage
\section{Introduction}
The existence of small neutrino masses \cite{nexp} and dark matter (DM) 
\cite{uobs} gives us important clues to consider  physics beyond the 
standard model (SM). It is interesting that various recent works clarify
a certain type of models can relate them closely. 
Such typical examples are neutrino models in which neutrino masses are
radiatively generated at TeV regions. There, this characteristic 
feature is caused by a symmetry which forbids the generation of 
the Dirac neutrino masses at tree-level and guarantees 
the stability of new neutral particles, simultaneously. 
A stable particle among them might be DM.
Since both the relic abundance of DM and the neutrino oscillation data 
severely constrain the relevant model parameters in general, 
the viability of such scenarios is expected to be checked 
through near future experiments at TeV or lower energy regions.  

A concrete model can be constructed on the basis of the 
radiative seesaw mechanism proposed in \cite{ma}. 
It is a very simple extension of the SM 
by an additional doublet scalar and three right-handed 
neutrinos only.\footnote{We call this new doublet the inert doublet
hereafter, although it has Yukawa couplings with neutrinos.} 
A $Z_2$ symmetry is introduced in the model such that the SM contents
have even parity and the new particles have odd parity. 
Then, it can forbid the generation of neutrino masses at tree-level 
and also guarantees the stability of the lightest $Z_2$ odd particle
which could be DM.
In this model, we can consider various possible scenarios depending on 
the spectrum of new particles which determines a DM candidate 
\cite{fcnc,flavor,warm}.\footnote{Supersymmetric extension has also been
considered in \cite{susy}.}

If the lightest right-handed neutrino is identified with the DM,
its neutrino Yukawa couplings are generally required to take large values.
In such a case, both the lepton flavor violating processes and the relic
abundance of DM can easily contradict each other \cite{fcnc}.
In order to evade this problem, we need to assume special 
flavor structure or introduce some new interaction at 
TeV regions \cite{flavor}. In the thermal leptogenesis due to the decay
of the right-handed neutrinos \cite{fy,leptg}, the washout of the generated 
lepton asymmetry is too effective to yield the required baryon number
asymmetry in this case . We need to consider non-thermal leptogenesis
\cite{nonth} or resonant leptogenesis \cite{tlept}.
   
On the other hand, if the lightest neutral component of the inert
doublet scalar is identified with DM, this problem can be
escaped.\footnote{The inert doublet DM has been studied in a lot of
papers \cite{inertdm}.} 
Since the scalar quartic couplings effectively cause their 
(co)annihilation, the neutrino Yukawa couplings could be irrelevant 
to the DM relic abundance \cite{ham}.
Thus, the lepton flavor violating processes do not impose substantial
constraints on the model in this case.
Thus, if we consider the leptogenesis there,
the washout of the generated lepton number asymmetry can be suppressed 
by making the neutrino Yukawa couplings small enough.
The model has been shown to explain the DM abundance and the baryon number 
asymmetry in a consistent way with the neutrino oscillation data
assuming the normal hierarchy in \cite{ks}.
Although the resonant effect is indispensable to enhance the $CP$
asymmetry sufficiently, the degeneracy required for the
right-handed neutrino masses is found to be rather mild there.
    
In this paper we extend the study in \cite{ks} to the 
inverted neutrino mass hierarchy.
We examine whether all the neutrino oscillation data and the baryon
number asymmetry can be consistently explained also in such a framework.
We also discuss how the generation mechanism of the neutrino masses 
in this model could be related to near future experiments 
for a direct DM search and also the neutrinoless double $\beta$ decay. 

The following parts are organized as follows. 
In section 2 we briefly introduce the model and discuss the realization 
of the inverted neutrino mass hierarchy. We also address the necessary
conditions to consider the thermal leptogenesis in this model. 
In section 3 we give the result of the numerical analysis of the baryon
number asymmetry. Taking account of this result, the relation 
between this neutrino mass generation mechanism and the phenomena at the low 
energy region such as the DM scattering with nucleus and 
the neutrinoless double $\beta$ decay is discussed. 
The conclusion of the paper is given in section 4.

\section{Inverted mass hierarchy in a radiative neutrino mass model}
We consider a simple extension of the standard model (SM) which 
is proposed for the neutrino mass generation at one-loop level \cite{ma}.
In this model, only three right-handed neutrinos $N_i$ and an inert 
doublet scalar $\eta$ are added to the SM as new ingredients. 
Although both $N_i$ and $\eta$ are supposed to have odd parity 
of an assumed $Z_2$ symmetry, all SM contents are assigned 
its even parity.
$Z_2$ invariant Yukawa couplings and scalar potential 
related to these new fields are summarized as
\begin{eqnarray}
-{\cal L}_Y&=&h_{\alpha i} \bar N_i\eta^\dagger\ell_\alpha
+h_{\alpha i}^\ast\bar\ell_\alpha\eta N_i+
\frac{M_i}{2}\bar N_iN_i^c 
+\frac{M_i}{2}\bar N_i^cN_i, \nonumber \\
V&=&m_\phi^2\phi^\dagger\phi+m_\eta^2\eta^\dagger\eta+
\lambda_1(\phi^\dagger\phi)^2+\lambda_2(\eta^\dagger\eta)^2
+\lambda_3(\phi^\dagger\phi)(\eta^\dagger\eta) \nonumber \\ 
&+&\lambda_4(\eta^\dagger\phi)(\phi^\dagger\eta) 
+\Big[\frac{\lambda_5}{2}(\phi^\dagger\eta)^2 + {\rm h.c.}\Big], 
\label{model}
\end{eqnarray}
where $\ell_\alpha$ is a left-handed lepton doublet and $\phi$ is 
an ordinary Higgs doublet with $|\langle\phi\rangle|=174$~GeV. 
Yukawa couplings are written by using 
the basis, under which both matrices of the right-handed neutrino masses 
and the Yukawa couplings of charged leptons are real and diagonal.
Since the new doublet scalar $\eta$ is assumed to have no vacuum 
expectation value,  the $Z_2$ symmetry is remained as the exact
symmetry. It forbids the neutrinos to have Yukawa interactions with
the ordinary Higgs scalar $\phi$. 
As a result, neutrino masses are not generated at
tree-level as found from eq.~(\ref{model}). 
The lightest field with odd parity of this $Z_2$ symmetry is stable and then
its thermal relics behave as DM in the Universe.
If the lightest neutral component of $\eta$ is identified with DM, 
it is found that its (co)annihilation caused by the scalar quartic 
couplings $\lambda_3$ and $\lambda_4$ can determine its relic 
abundance \cite{ham,ks}.
In this case, since the DM abundance gives no constraint on the neutrino 
Yukawa couplings, the model can be easily consistent with other 
phenomenological constraints such as the ones caused by the 
lepton flavor violating processes. 
We follow this scenario in this paper, and $\eta_R$ is assumed to be DM
which requires $\lambda_5<0$ and $\lambda_4<0$ \cite{ks}.

Neutrino masses are generated through one-loop diagrams with 
the contribution of the $Z_2$ odd particles. They can be expressed as
\begin{equation}
{\cal M}^\nu_{\alpha\beta}=\sum_{k=1}^3h_{\alpha k}h_{\beta k}\Lambda_k.
\label{nmass1}
\end{equation}
Scales for the neutrino masses are considered to be 
fixed by $\Lambda_k$, which is 
defined as
\begin{equation}
\Lambda_k=\frac{\lambda_5\langle\phi\rangle^2}
{8\pi^2M_k}\frac{M_k^2}{M_\eta^2-M_k^2}
\left(1+\frac{M_k^2}{M_\eta^2-M_k^2}\ln\frac{M_k^2}{M_\eta^2}\right),
\label{nmass2}
\end{equation}
where $M_\eta^2=m_\eta^2+(\lambda_3+\lambda_4)
\langle\phi\rangle^2$.\footnote{We note that the $\lambda_5$ contribution
in this formula for $M_\eta^2$ is neglected, since $\lambda_5$ is assumed 
to be sufficiently small.}
This shows that neutrino Yukawa couplings could have rather 
large values even for the light right-handed neutrinos with the mass 
of $O(1)$~TeV as long as $|\lambda_5|$ takes a small value. 

Now we consider the realization of the inverted hierarchy in this
mass formula. For this purpose, we may start the study from the 
neutrino mass matrix which brings 
the tri-bimaximal mixing.\footnote{In this paper we do not assume any
flavor symmetry to realize this structure. Thus, the quantum corrections 
could change it. However, since the assumed neutrino Yukawa couplings are
very small, the zero texture of these neutrino Yukawa couplings are 
expected to be kept in good accuracy after taking account of the
quantum effects through the renormalization group equations. Thus, 
if the values of nonzero neutrino Yukawa couplings at high
energy scale are set suitably, the results obtained in this study could 
be reproduced.
The detailed analysis is beyond the scope of the present study 
and it will be given elsewhere.}
Since the recent experiments suggest nonzero $\theta_{13}$ \cite{t13}, 
it can be just an approximation for the realistic mixing.
However, we use it as a convenient starting point of the study 
and introduce suitable modifications to it so as to realize 
a favorable value for $\theta_{13}$.  Taking this strategy, 
we assume that the neutrino Yukawa couplings have the following
flavor structure \cite{flavor}:
\begin{equation}
h_{e1}=-2h_{\mu 1}=2h_{\tau 1}=2h_1; \quad
h_{e2}=h_{\mu 2}=-h_{\tau 2}= h_2; \quad
h_{e3}=0,~ h_{\mu 3}=h_{\tau 3}= h_3. 
\label{yukawa}
\end{equation}
This flavor structure for the neutrino Yukawa couplings induces 
a following simple form for the neutrino mass matrix:
\begin{equation}
{\cal M}^\nu=
\left(
\begin{array}{ccc}
4 & -2 & 2\\ -2 & 1 & -1 \\ 2 & -1 & 1 \\ 
\end{array}\right)h_1^2\Lambda_1
+\left(
\begin{array}{ccc}
1 & 1 & -1\\ 1 & 1 & -1 \\ -1 & -1 & 1 \\ 
\end{array}\right)h_2^2\Lambda_2
+\left(
\begin{array}{ccc}
0 & 0 & 0\\ 0 & 1 & 1 \\ 0 & 1 & 1 \\ 
\end{array}\right)h_3^2\Lambda_3.
\label{nmatrix}
\end{equation} 
As easily found, this matrix can be diagonalized as
$U_{\rm PMNS}^T{\cal M}^\nu U_{\rm PMNS}={\rm diag}(m_1,~m_2,~m_3)$ 
by using the tri-bimaximal PMNS matrix 
\begin{equation}
U_{\rm PMNS}=\left(\begin{array}{ccc}
\frac{2}{\sqrt 6} & \frac{1}{\sqrt 3} & 0\\
 \frac{-1}{\sqrt 6} & \frac{1}{\sqrt 3} & \frac{1}{\sqrt 2}\\
\frac{1}{\sqrt 6} & \frac{-1}{\sqrt 3} & \frac{1}{\sqrt 2}\\
\end{array}\right)
\left(\begin{array}{ccc}
e^{i\alpha_1} &0 & 0\\
0 & e^{i\alpha_2} & 0 \\
0 & 0 & 1 \\
\end{array}\right). 
\end{equation}
The mass eigenvalues are expressed as
\begin{equation}
m_1=6|h_1^2\Lambda_1|, \qquad m_2=3|h_2^2\Lambda_2|, 
\qquad m_3=2|h_3^2\Lambda_3|.
\label{nmeig}
\end{equation}
Majorana phases $\alpha_{1,2}$ are determined by the phases
$\varphi_i={\rm arg}(h_i)$ as
\begin{equation}
\alpha_1=\varphi_1-\varphi_3, \qquad
\alpha_2=\varphi_2-\varphi_3.
\label{cpphase}
\end{equation}
where $\langle\phi\rangle$ is assumed to be real and positive.

Here we impose the neutrino oscillation data on the model.
They are well-known to be explained by the two types of mass hierarchy
which are called the normal hierarchy and the inverted hierarchy.
In the present model, these possibilities are realized in the following way.
In the former case, the mass eigenvalues should satisfy
\begin{equation}
m_3^2-m_1^2=\Delta m^2_{\rm atm}, \qquad m_2^2-m_1^2=\Delta m^2_{\rm sol},
\end{equation} 
where $\Delta m^2_{\rm atm}$ and $\Delta m^2_{\rm sol}$ stand for 
squared mass differences required by the neutrino oscillation analysis
for atmospheric and solar neutrinos, respectively \cite{nexp,prg,bestf}.
This case has been investigated from various phenomenological 
points of view \cite{fcnc,flavor,warm,ks}.
Through this study, it is shown that all the neutrino masses, mixing and
the DM abundance are successfully explained. 
However, the generation of the baryon number asymmetry due to 
thermal leptogenesis is difficult unless the resonance effect 
is taken into account. 
On the other hand, in the latter case, the mass eigenvalues should satisfy
\begin{equation}
\Delta m_{13}^2\equiv m_1^2-m_3^2=\Delta m^2_{\rm atm}, 
\qquad \Delta m_{21}^2\equiv m_2^2-m_1^2=\Delta m^2_{\rm sol}.
\label{inv}
\end{equation} 
Although this case is expected to have interesting features which are different 
from those of the normal hierarchy, it seems not to have been studied 
well in this model yet. 
We focus our attention on this case in the remaining part.

Before considering the concrete realization of the inverted hierarchy, 
we first address necessary conditions which should be satisfied for the
thermal leptogenesis brought through the decay of the lightest 
right-handed neutrino in this framework.
The lightest neutral component of $\eta$ which has the mass in the 
TeV region is known to satisfy the required DM relic abundance 
as long as the quartic couplings $\lambda_3$ or $\lambda_4$ have 
the sufficient magnitude \cite{ham,ks}.
In that case, its relic abundance is determined only by these couplings.
Since neutrino Yukawa couplings play no role in this determination, 
the relic DM abundance has no relation with the neutrino mass
eigenvalues as found from eqs.~(\ref{nmass1}) and (\ref{nmass2}) as in
the normal hierarchy case.
We follow this scenario and assume that $M_\eta< M_{1,2,3}$ is satisfied.
In this case, the present direct searches of DM \cite{direct1,direct2} 
impose the constraint on $\lambda_5$ and $\lambda_3+\lambda_4$.

The bound for $|\lambda_5|$ appears from the inelastic scattering 
between $\eta_{R,I}^0$ and a nucleus through $Z^0$ exchange which could 
contribute to direct search experiments.
The differential scattering rate for recoil nucleus energy $E_R$
crucially depends on the DM velocity distribution and it has a
relation such as \cite{susydm}
\begin{equation}
\frac{dR}{dE_R} \propto \sigma_n^0
\int_{v_{\rm min}}^{v_{\rm esc}}d^3v~ \frac{f(v)}{v},
\end{equation}
where $f(v)$ is the DM velocity distribution function and $\sigma_n^0$
is the DM-nucleon scattering cross section with zero momentum transfer.
Since $\eta_R^0$ and $\eta_I^0$ have mass difference $\delta$, 
the scattering occurs only for the DM velocity larger than \cite{vmin}
\begin{equation} 
v_{\rm min}=\frac{1}{\sqrt{2m_NE_R}}\left(\frac{E_Rm_N}{m_r}
+\delta\right),
\end{equation}
where $m_N$ is the mass of a target nucleus and $m_r$ is 
the reduced mass of DM and the nucleus.
Upper bound of the DM velocity is determined as the escape velocity from
the galaxy and it is estimated as $v_{\rm esc}= 544$~km/s.
In Fig.~1, we plot the present Xenon100 bound on $\sigma_n^0$ as a
function of $\delta$. The lines representing this bound are found 
to have an end point for large values of $\delta$. 
This is caused since $v_{\rm esc}<v_{\rm min}$ occurs there. 
In that case, the inelastic scattering is kinematically forbidden 
due to the existence of the upper bound of DM velocity $v_{\rm esc}$. 
Since $\sigma_n^0$ for this inelastic scattering is estimated
as $\sigma_n^0\simeq \frac{1}{2\pi}G_F^2m_n^2\sim 7.4\times 
10^{-39}~{\rm cm}^2$ and $\delta\equiv m_{\eta_I}-m_{\eta_R}\simeq 
\frac{|\lambda_5|\langle\phi\rangle^2}{m_{\eta_R}}$,
we find that $|\lambda_5|$ is constrained from Fig.~1 as \cite{l5}
\begin{equation}
|\lambda_5|\simeq \frac{M_\eta \delta}{\langle\phi\rangle^2} 
~~{^>_\sim}~~6.7\times 10^{-6}
\left(\frac{M_\eta}{1~{\rm TeV}}\right)
\left(\frac{\delta}{200~{\rm keV}}\right).
\label{direct1}
\end{equation}
This bound affects the neutrino Yukawa couplings through the 
neutrino oscillation data if we recall the present 
neutrino mass generation mechanism represented by 
eqs.~(\ref{nmass1}) and (\ref{nmass2}).

\input epsf
\begin{figure}[t]
\begin{center}
\epsfxsize=8cm
\leavevmode
\epsfbox{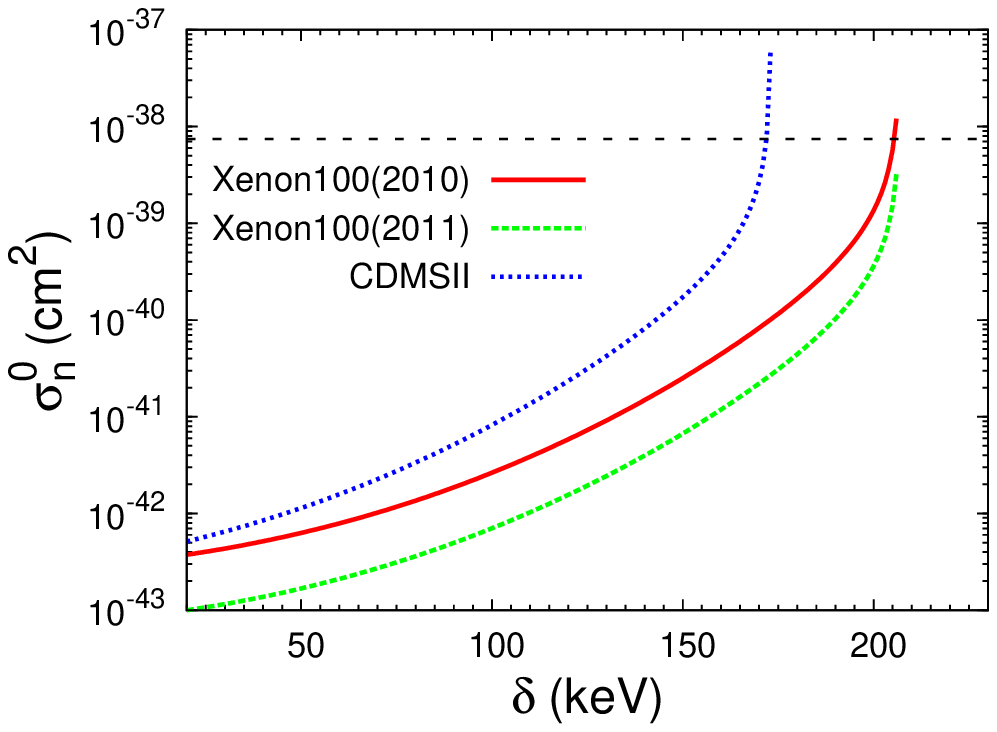}
\end{center}
\vspace*{-3mm}

{\footnotesize {\bf Fig.~1}~~Direct detection constraint for the mass
 difference $\delta$ between $\eta_R$ and $\eta_I$.  
The bound of the DM-nucleon cross section obtained in each experiment is 
drawn. In this plot we use $E_R=40$~keV, as an example. }   
\end{figure}

The elastic scattering due to the Higgs exchange can also be a target of
the direct search. The present direct search results impose the 
constraint on the value of $\lambda_3+\lambda_4$, 
since the DM-nucleon scattering cross section 
for this process is estimated as
\begin{equation}
\sigma_n^0=\frac{(\lambda_3+\lambda_4+\lambda_5)^2}{8\pi}
\frac{m_n^2f_n^2}{m_{\eta_R}^2m_h^4}. 
\end{equation}  
The present bound obtained by Xenon100 gives the constraint such as
\begin{equation}
|\lambda_3+\lambda_4+\lambda_5|~{^<_\sim}~1.8
\left(\frac{m_{\eta_R}}{1~{\rm TeV}}\right),
\label{direct2}
\end{equation}
where we use $m_h=125$~GeV and $f_n=1/3$. 
This bound could be intimately related to the relic density of 
$\eta_R$ \cite{ks}.
It may be useful to note that this condition can be easily satisfied in
the case $\lambda_3\lambda_4<0$ even for $|\lambda_{3,4}|=O(1)$, which is
favored to reduce effectively the $\eta_R$ relic density 
to the required value.
This point will be discussed in the relation to the DM detection at the
Xenon1T experiment in the latter part again.
  
On the other hand, thermal leptogenesis requires 
that the decay of the lightest right-handed neutrino $N_i$ should occur 
in the out-of-thermal equilibrium.
This imposes the condition such that $H>\Gamma_{N_i}^D$ has to be 
satisfied at $T~{^<_\sim}~ M_i$, where $H$ and $\Gamma_{N_i}^D$ are 
the Hubble parameter and the decay width of $N_i$, respectively. 
From this condition, we find that the Yukawa coupling of $N_i$ 
should be small enough as 
\begin{equation}
|h_i| < O(10^{-8}) \left(\frac{M_i}{1~{\rm TeV}}\right)^{1/2}.
\label{h1}
\end{equation}
Here we note that small $|h_i|$ guarantees that $N_i$ is irrelevant 
to the determination of the neutrino masses and the mixing angles.
If we apply this condition to eqs.~(\ref{nmeig}) and (\ref{inv}), 
we find that the lightest right-handed neutrino should be $N_3$ in the
present case.
This means that both $|h_3|<O(10^{-8})$ and $M_3<M_{1,2}$ should be satisfied.
Since the resonant leptogenesis is considered to be crucial 
for the generation of the sufficient baryon number asymmetry as
in the normal hierarchy case, 
we can suppose two scenarios defined by the following spectrum of the
right-handed neutrinos here:
\begin{equation}
{\rm (i)}~ M_3~{^<_\sim}~ M_1<M_2, \qquad {\rm (ii)}~ M_3~{^<_\sim}~ M_2<M_1,
\label{degeneracy}
\end{equation}   
where the first inequality represents that these masses are almost 
degenerate each other in each 
case.\footnote{It is possible to consider the situation such
that three right-handed neutrinos are all degenerate. However, we do not
consider this case here since the result can be estimated by using 
the results of the two cases.}

We present a brief comment on the degenerate masses of 
the right-handed neutrinos.\footnote{Other scenarios for tiny mass
splittings of the right-handed neutrino can be found in \cite{split}.} 
They could be obtained without disturbing the flavor structure 
of ${\cal L}_Y$ in eq.~(\ref{model}), if their mass terms take a form such as
\begin{equation}
{\cal L_M}=\frac{M}{2}\left(\bar N_3N_3^c +\bar N_iN_i^c\right) 
+ \frac{M_j}{2}\bar N_jN_j^c +
\frac{\Delta_i}{2}M\bar N_3N_i^c +{\rm h.c.},
\label{rmass} 
\end{equation} 
where $\Delta_i\ll 1$ is assumed. 
It is useful to note that this structure for the right-handed neutrino 
masses could be realized by supposing the model with the fifth dimension.
If we assume that the right-handed neutrinos localize around the different 
fifth dimensional points and other fields are distributed throughout the
fifth dimensional direction, the neutrino sector in the effective 
4-dimensional model obtained after integrating out the fifth 
coordinate can have the feature described by ${\cal L}_Y$ and 
${\cal L}_M$.\footnote{Although the high degeneracy of the
right-handed neutrino masses might be explained in this way, 
the required values for Yukawa couplings in ${\cal L}_Y$ should be 
just assumed in this framework.}
The small mass mixing between $N_3$ and $N_i$ is explained by the small
overlapping of their wave function in the direction of the 
fifth dimension.  
In the case described by eq.~(\ref{rmass}), the mass eigenstates in the
mixed sector can be identified with $N_3$ and $N_i$ 
to a good accuracy and their 
mass eigenvalues are $M_3=M(1-\frac{\Delta_i}{2})$ 
and $M_i=M(1+\frac{\Delta_i}{2})$. Using the mass eigenvalues, 
$\Delta_i$ is expressed as $\Delta_i=\frac{M_i}{M_3}-1$.
If we take $i=1$ and $j=2$, for example, the case (i) is realized.  
 
\begin{figure}[t]
\begin{center}
\epsfxsize=6cm
\leavevmode
\epsfbox{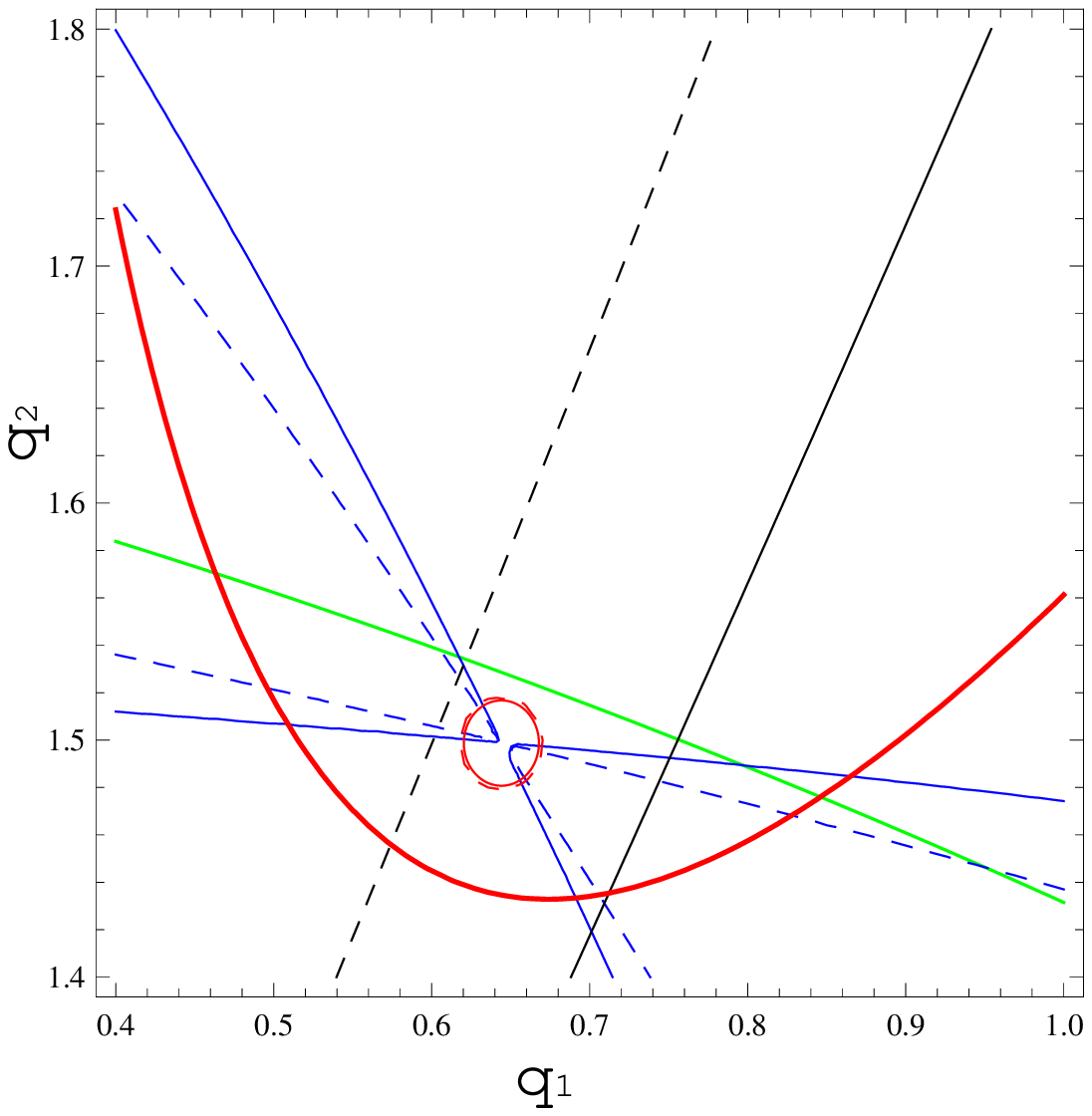}
\hspace*{10mm}
\epsfxsize=6cm
\leavevmode
\epsfbox{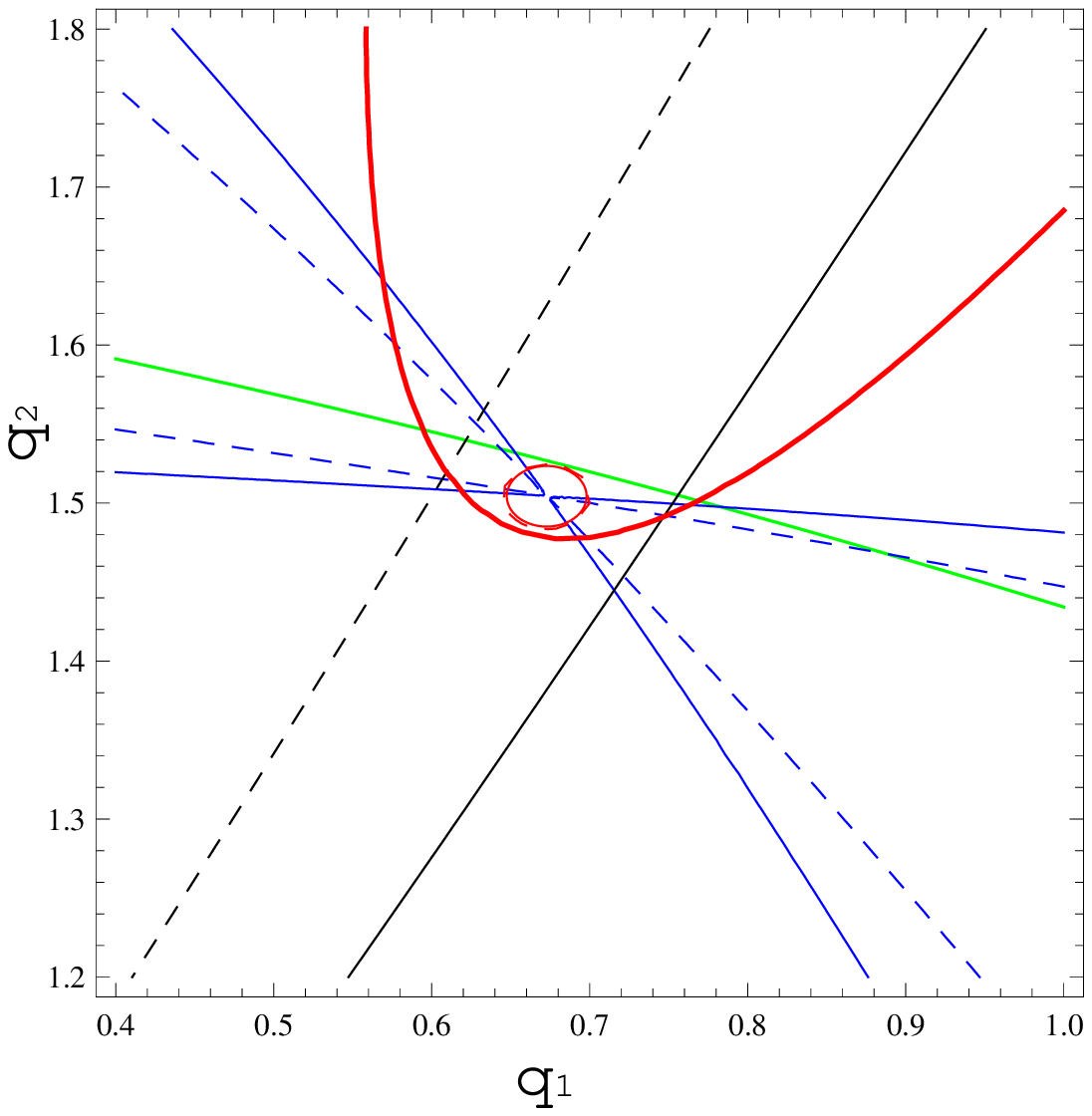}
\end{center}
\vspace*{-3mm}

{\footnotesize {\bf Fig.~2}~~Allowed region in the $(q_1,q_2)$ plane
by the neutrino oscillation data. Left and right panels correspond to
 the cases (i) and (ii) shown in Table 1, respectively.
Each contour in both panels represents neutrino oscillation 
parameters $\Delta m^2_{13}$ (thick red lines), $\Delta m_{21}^2$
 (thin red lines), $\sin^22\theta_{23}$ (green lines), 
$\sin^22\theta_{12}$ (blue lines), 
and $\sin^22\theta_{13}$ (black lines) for the 2$\sigma$ values 
given in \cite{bestf}. For each of the parameters, the upper and lower bounds 
are expressed by the dashed and solid lines, respectively.}   
\end{figure}

Now we fix the framework to realize the inverted hierarchy 
which can explain the neutrino oscillation data. 
Recent experiments show that the mixing angle $\theta_{13}$ is not
zero but has a rather large value \cite{t13}. 
In order to make the explanation of these data possible in the model, 
the flavor structure (\ref{yukawa}) has to 
be changed. Although there are a lot of way to introduce the 
modification for it, we take a simple scheme such that 
eq.~(\ref{yukawa}) is deformed by new real free parameters 
$p_{1,2}$ and $q_{1,2}$ as   
\begin{equation}
h_{e1}/p_1=-2h_{\mu 1}/q_1=2h_{\tau 1}=2h_1; \quad
h_{e2}/p_2=h_{\mu 2}/q_2=-h_{\tau 2}= h_2. 
\label{myukawa}
\end{equation}
Since the smallness of $|h_3|$ guarantees that $N_3$ is irrelevant to the
neutrino masses and mixing, the neutrino mass matrix can be written as
\begin{equation}
{\cal M}^\nu\simeq
\left(
\begin{array}{ccc}
4p_1^2 & -2p_1q_1 & 2p_1\\ -2p_1q_1 & q_1^2 & -q_1 \\ 2p_1 & -q_1 & 1 \\ 
\end{array}\right)h_1^2\Lambda_1
+\left(
\begin{array}{ccc}
p_2^2 & p_2q_2 & -p_2\\ p_2q_2 & q_2^2 & -q_2 \\ -p_2 & -q_2 & 1 \\ 
\end{array}\right)h_2^2\Lambda_2.
\label{nmatrix1}
\end{equation}  

We numerically diagonalize this matrix to find the mass eigenvalues
and the mixing angles. The model 
parameters are determined by comparing these results with the 
neutrino oscillation data.
In Fig.~2, we plot contours in the $(q_1,~q_2)$ plane for 2$\sigma$
values of neutrino oscillation parameters given in \cite{bestf}. 
In each panel of this figure, a part of parameters is commonly fixed as 
$M_\eta=1$~TeV, $M_3=2$~TeV and $|\lambda_5|=10^{-4}$. The values used
for other parameters $M_{1,2}$, $|h_{1,2}|$ and $p_{1,2}$ are shown 
in Table 1.
In this figure, we find that the allowed regions are displayed as 
four sectors on the small circle drawn by the thin red lines, which are
sandwiched by the blue dashed and solid lines.
Values of $(q_1, q_2)$ shown in Table 1 are contained 
in these regions.\footnote{In this analysis, we find the
 solutions by varying $(q_1,q_2)$ only, for simplicity. If we vary other
 parameters to find solutions simultaneously, the tuning of 
$(q_1,q_2)$ required here is expected to be much mild.}
The predicted value of $\sin^22\theta_{13}$ at this point 
is also given in this table.   
This figure shows that the mass matrix derived from the flavor structure
(\ref{myukawa}) can explain all the neutrino oscillation data,
the DM relic abundance and its phenomenology
consistently as long as the relevant parameters are fixed suitably.

\begin{figure}[t]
\begin{center}
\begin{tabular}{ccccccccc}\hline
&$M_1$ & $M_2$ &$10^4|h_1|$ & $10^4|h_2|$ & $(p_1,p_2)$ &  $(q_1, q_2)$ &
 $\sin^22\theta_{13}$ & $m_{ee}$({\rm eV})\\ \hline
(i)& 2 & 8 & 5.8 & 9.0 & (1.17,0.84) &  (0.630,1.515) & 0.11 &
0.0195-0.0486\\
(ii)& 8 & 2 & 6.7 & 7.5 & (1.17,0.86) & (0.656,1.521) & 0.10 &
0.0156-0.0473\\
\hline
\end{tabular}
\end{center}
\vspace*{3mm}

{\footnotesize Table~1~  Typical parameters used in this analysis. Other
 parameters are fixed as $|h_3|=3\times 10^{-8}$, $|\lambda_5|=10^{-4}$,
$M_3=2$~TeV, and $M_\eta=1$~TeV. }
\end{figure}

As shown in these analysis, small neutrino Yukawa couplings 
$|h_{1,2}|$ of $O(10^{-3})$ can explain the neutrino oscillation data 
even for the TeV scale values of $M_{1,2}$ and $M_\eta$ as long as we assume 
the values of $O(10^{-4})$ for $|\lambda_5|$. 
Although both the lepton flavor violating processes such as 
$\mu\rightarrow e\gamma$ and the anomalous magnetic momentum of a muon 
are induced through one-loop diagrams with $Z_2$ odd particles in the
internal lines \cite{mr}, these contributions could be largely 
suppressed due to these small neutrino Yukawa couplings. 
In fact, for the parameters used here, we find that these quantities are
negligibly small as follows,
\begin{eqnarray} 
&&{\rm Br}(\mu\rightarrow e\gamma)=
\frac{3\alpha}{64\pi(G_FM_\eta^2)^2}\left[-2p_1h_1^2
F_2\left(\frac{M_1}{M_\eta}\right)
+p_2h_2^2F_2\left(\frac{M_2}{M_\eta}\right)\right]^2=O(10^{-25\sim -24}), 
\nonumber \\
&&{\rm Br}(\tau\rightarrow \mu\gamma)=
\frac{3\alpha}{64\pi(G_FM_\eta^2)^2}\left[q_1h_1^2
F_2\left(\frac{M_1}{M_\eta}\right)
+q_2h_2^2F_2\left(\frac{M_2}{M_\eta}\right)\right]^2=O(10^{-24\sim -23}), 
\nonumber \\ 
&&\delta a_\mu=\frac{1}{(4\pi)^2}\frac{m_\mu^2}{M_\eta^2}
\left[h_1^2F_2\left(\frac{M_1}{M_\eta}\right)+
h_2^2F_2\left(\frac{M_2}{M_\eta}\right)\right]=O(10^{-18}), 
\end{eqnarray}
where $F_2(x)=(1-6x^2+3x^4+2x^6-6x^4\ln x^2)/6(1-x^2)^4$.
The present bounds for these lepton flavor violating processes \cite{lv}
do not impose any constraints on the model.
Since the contribution of the new particles to the muon $g-2$ is 
insufficient to account for the experimental value
which deviates from the SM prediction \cite{g-2},
some additional extension of the model is required for that explanation.
If we make $|\lambda_5|$ smaller, these values become larger due to the
larger neutrino Yukawa couplings. However, as long as 
the value of $|\lambda_5|$ takes a value in the region given in 
eq.(\ref{direct1}) which is imposed by the DM direct search, 
these processes can never be targets of future experiments to
examine the model unfortunately.

\section{Baryon number asymmetry and low energy phenomena}
\subsection{Resonant leptogenesis}
We consider the thermal leptogenesis due to the out-of-equilibrium decay
of the right-handed neutrino $N_3$.\footnote{The leptogenesis for the
high mass right-handed neutrinos is possible also in this radiative mass
model. In fact, such a possibility has been studied in \cite{ks} for 
the normal hierarchy and a similar result is expected for the
inverted hierarchy. We are interested in the features of the TeV
scale model here.}
In the previous part, we have considered two possible patterns 
for the masses of the right-handed neutrinos, which are 
shown in eq.~(\ref{degeneracy}).
Through this analysis, it has been shown that the model can 
explain all the neutrino oscillation data in the suitable parameter 
regions. These typical examples are shown in Fig.~2 and Table 1. 
Although the mass of $N_3$ seems to be 
too small to realize the sufficient $CP$ asymmetry associated with 
this decay process for the generation of the required baryon number 
asymmetry, it could be enhanced by the resonance effect as known in the 
ordinary resonant leptogenesis \cite{resonant}. 
The dominant contribution to the $CP$ asymmetry is expected 
to be caused by the interference between the tree diagram and 
the one-loop self-energy diagram as usual.
The $CP$ asymmetry $\varepsilon$ in the $N_3$ decay 
is expressed as \cite{resonant}
\begin{equation}
\varepsilon=\sum_{i=1,2}\frac{{\rm Im}
(h^\dagger h)^2_{i3}}{(h^\dagger h)_{33}
(h^\dagger h)_{ii}}\frac{(M_3^2-M_i^2)M_3\Gamma_{N_i}}
{(M_3^2-M_i^2)^2+M_3^2\Gamma_{N_i}^2},
\end{equation}
where $\Gamma_{N_i}=\sum_{\alpha}\frac{|h_{\alpha i}|^2}
{8\pi}M_i(1-\frac{M_\eta^2}{M_i^2})^2$.

The dominant part of $\varepsilon$ in each case shown in Table 1 is 
obtained by substituting 
$i=1$ for (i) and $i=2$ for (ii). Thus, if we use the flavor 
structure of neutrino Yukawa couplings assumed in 
eq.~(\ref{myukawa}), the $CP$ asymmetry $\varepsilon$ in each case 
can be expressed as  
\begin{eqnarray}
{\rm (i)}&&\varepsilon\simeq 
-\frac{(-1+q_1)^2}{2(4p_1^2+1 +q_1^2)}
\frac{2\Delta_1\tilde\Gamma_{N_1}}
{4\Delta_1^2+\tilde\Gamma_{N_1}^2}\sin
2(\varphi_3-\varphi_1),\nonumber \\
{\rm (ii)}&&\varepsilon\simeq 
-\frac{(1-q_2)^2}{2(p_2^2+1+q_2^2)}
\frac{2\Delta_2\tilde\Gamma_{N_2}}
{4\Delta_2^2+\tilde\Gamma_{N_2}^2}\sin
2(\varphi_3-\varphi_2),
\label{cp}
\end{eqnarray}
where $\Delta_i=\frac{M_i}{M_3}-1$ and 
$\tilde\Gamma_{N_i}=\frac{\Gamma_{N_i}}{M_i}$. 
These formulas show that the $CP$ asymmetry $\varepsilon$ 
could have large values for the case with
$\Delta_i=O(\tilde\Gamma_{N_i})$ 
as long as both $q_{1,2}\not=1$ and $\sin 2(\varphi_3-\varphi_{1,2})\not=0$ 
are satisfied. As easily found,
the $CP$ phases appeared in eq.~(\ref{cp}) have no relation with the $CP$ phase
which controls the value of effective neutrino mass $m_{ee}$ 
for the neutrinoless double $\beta$ decay in eq.~(\ref{beta}).

On the other hand, the washout of the generated lepton number asymmetry 
could be brought by both the lepton number violating 2-2 scattering 
such as $\eta\eta\rightarrow\ell_\alpha\ell_\beta$ and
$\eta^\dagger\ell_\alpha\rightarrow\eta\bar\ell_\beta$ 
and also the inverse decay of $N_i$. 
However, if the relevant Yukawa couplings are small enough,
these processes could be nearly freezed out before the temperature 
of the thermal plasma decreases to $T~{^<_\sim}~M_3$.
Thus, the washout of the generated lepton number asymmetry is expected
to be suppressed sufficiently. 
In order to examine this quantitatively, we numerically solve the 
coupled Boltzmann equations for the number density of $N_{3}$ and 
the lepton number asymmetry. 
We introduce these number densities in the co-moving volume 
as $Y_{N_{3}}=\frac{n_{N_{3}}}{s}$ 
and $Y_L=\frac{n_\ell- n_{\bar \ell}}{s}$ by using the entropy density $s$. 
The Boltzmann equations for these are written as\footnote{The $\Delta
L=2$ scattering reaction densities $\gamma_N^{(2)}$ and
$\gamma_N^{(13)}$ involve the interference terms for the right-handed
neutrinos with tiny mass splittings. Although they are suggested to 
play a crucial role in \cite{dl}, its effect might be suppressed due to
very small neutrino Yukawa couplings (\ref{h1}) of the lightest 
right-handed neutrino $N_3$. }
\begin{eqnarray}
&&\frac{dY_{N_{3}}}{dz}=-\frac{z}{sH(M_3)}
\left(\frac{Y_{N_{3}}}{Y_{N_{3}}^{\rm eq}}-1\right)\left\{
\gamma_D^{N_3}+\sum_{i=1,2}\left(\gamma_{{N_3}{N_i}}^{(2)}+
\gamma_{{N_3}{N_i}}^{(3)}\right)\right\}, \nonumber \\
&&\frac{dY_L}{dz}=\frac{z}{sH(M_3)}\left\{
\varepsilon\left(\frac{Y_{N_{3}}}{Y_{N_{3}}^{\rm eq}}-1\right)\gamma_D^{N_3}
-\frac{2Y_L}{Y_\ell^{\rm eq}}\left(\sum_{i=1,2}\frac{\gamma_D^{N_i}}{4}
+\gamma_N^{(2)} +\gamma_N^{(13)}\right)\right\}, 
\label{bqn}
\end{eqnarray}
where $z=\frac{M_3}{T}$ and $H(M_3)=1.66g_\ast^{1/2}\frac{M_3^2}{m_{\rm pl}}$.
The equilibrium values for these are expressed as 
$Y_{N_{3}}^{\rm eq}(z)=\frac{45}{2\pi^4g_\ast}z^2K_2(z)$
and $Y_\ell^{\rm eq}=\frac{45}{\pi^4g_\ast}$, where $K_2(z)$ is the
modified Bessel function of the second kind.
In these equations we omit several processes whose contributions 
are considered to be negligible compared with others.
We should note that the inverse decay to $N_{1,2}$ induced by the Yukawa
couplings $h_{1,2}$ could cause a large contribution to the washout of
the lepton number asymmetry since $M_i\simeq M_3$ is satisfied for $i=1$
in case (i) and for $i=2$ in case (ii).\footnote{In the analysis of the
resonant leptogenesis in \cite{ks}, the inverse decay has not been taken
into account. As the result, the required mass degeneracy for the 
right-handed neutrinos is 
underestimated by one order of magnitude there. In that case, 
however, it is still milder than the usual case.}
In the Appendix, we present the relevant reaction density $\gamma$ 
contained in these equations. 

\begin{figure}[t]
\begin{center}
\epsfxsize=7cm
\leavevmode
\epsfbox{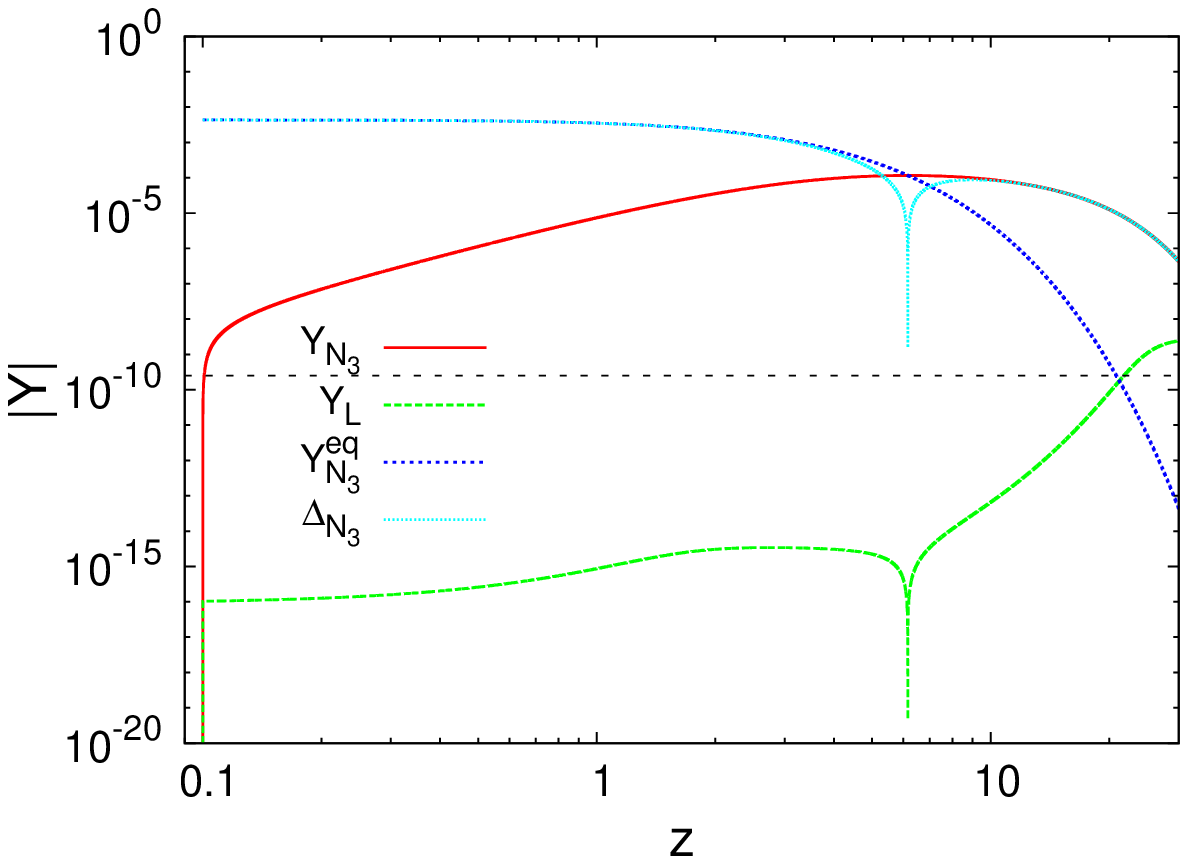} 
\epsfxsize=7cm
\leavevmode
\hspace*{10mm}
\epsfbox{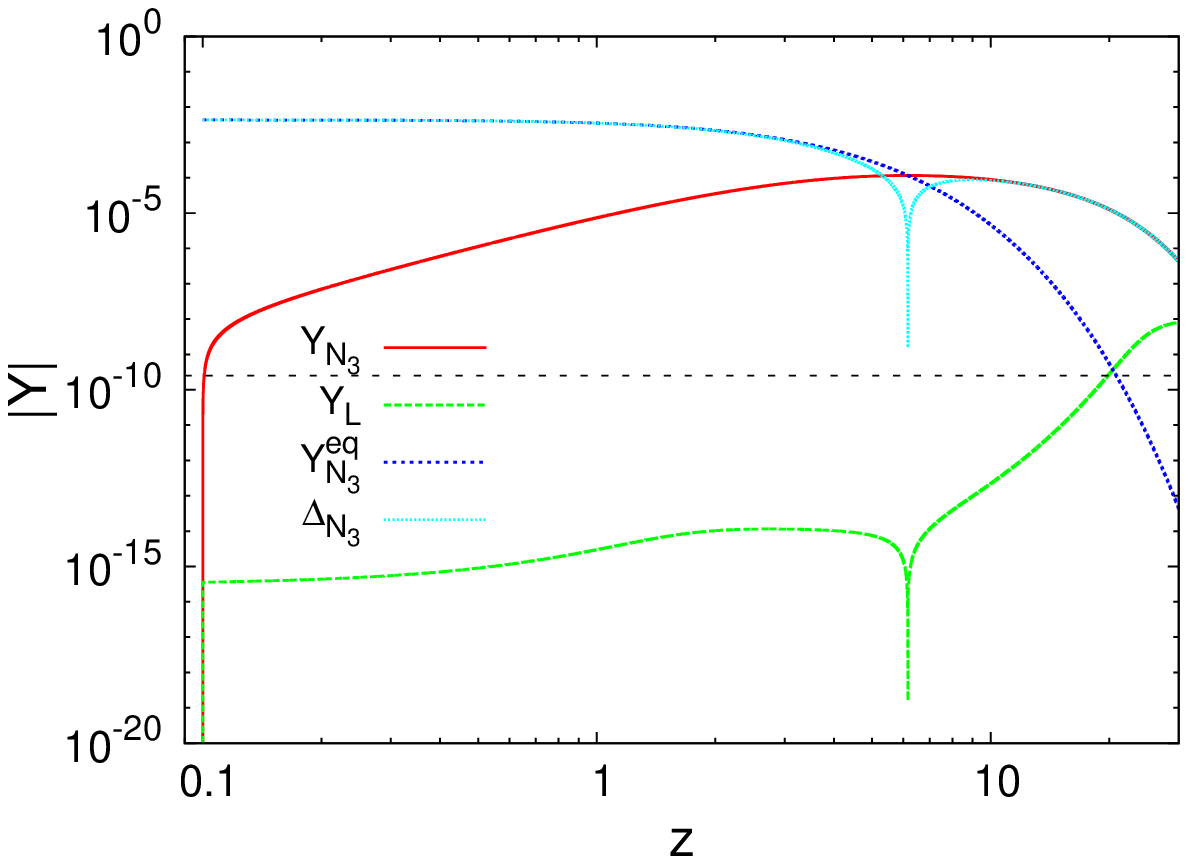}\\
\epsfxsize=7cm
\leavevmode
\epsfbox{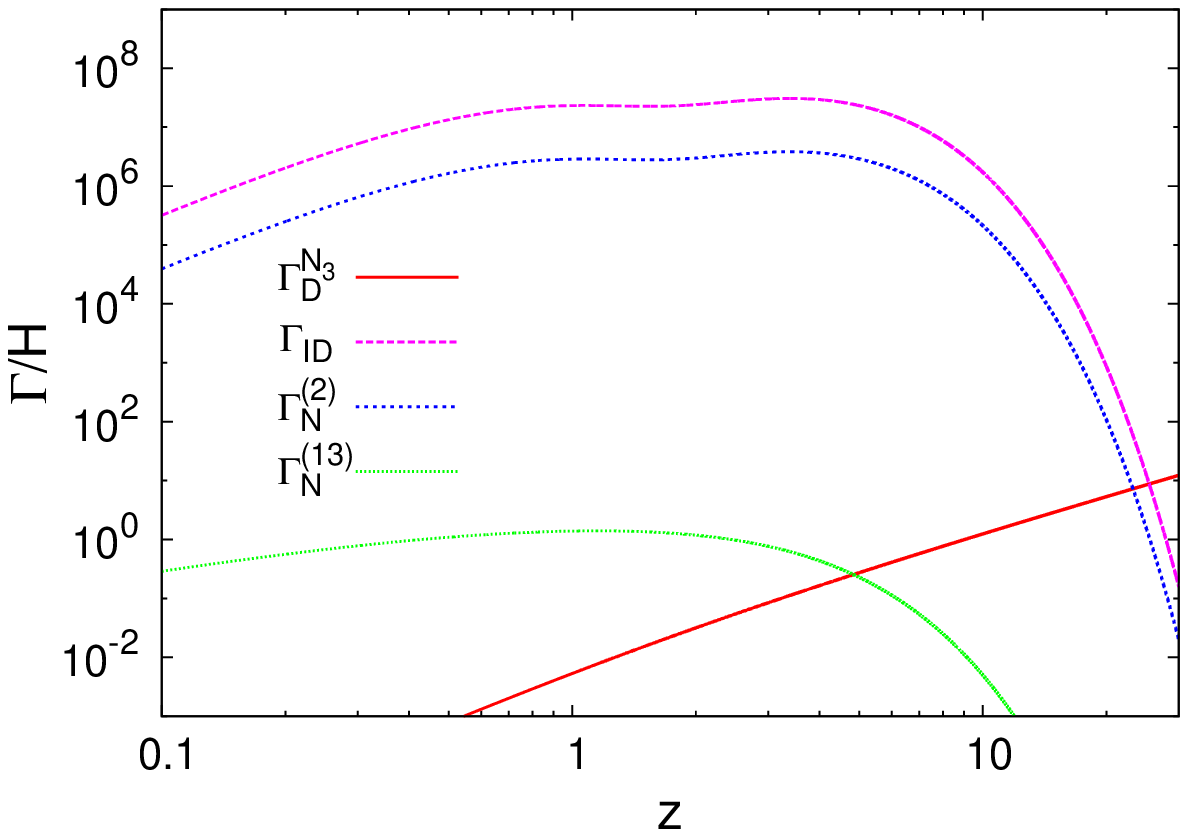} 
\epsfxsize=7cm
\leavevmode
\hspace*{10mm}
\epsfbox{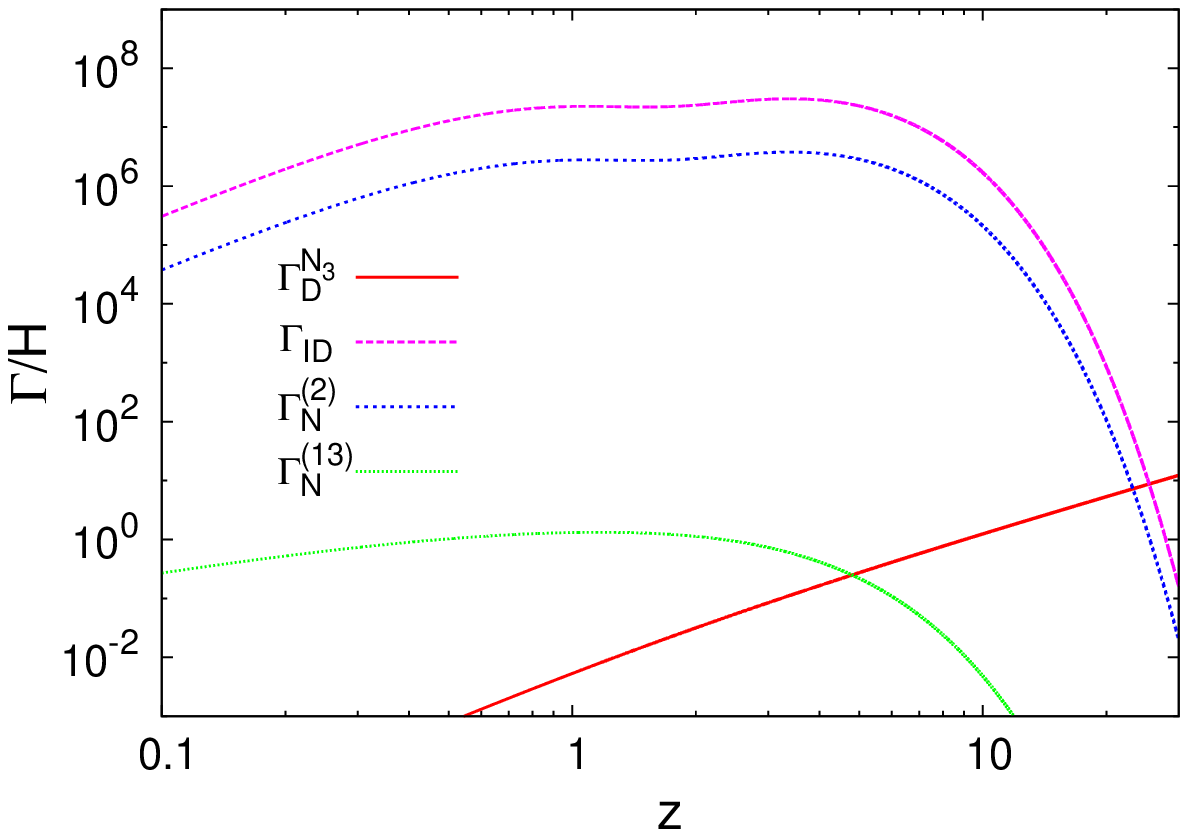}\\
\end{center}
\vspace*{-3mm}
{\footnotesize {\bf Fig.~3}~~ The upper left and right panels show 
the evolution of the lepton number asymmetry $|Y_L|$ and the $N_3$ 
number density $|Y_{N_3}|$ for the cases (i) and (ii) given in Table 1
with $\Delta_i=10^{-6.5}$.
The lower panels show the ratio of the reaction rate of the lepton
 number violating processes to the Hubble parameter in each 
corresponding case. }
\end{figure}

The baryon number asymmetry $Y_B(\equiv\frac{n_b-n_{\bar b}}{s})$ 
is transformed from the generated lepton number asymmetry through 
the sphaleron process.   
It is estimated by using the solution of the coupled equations in 
eq.~(\ref{bqn}) as
\begin{equation}
Y_B=-\frac{8}{23}Y_L(z_{\rm EW}),
\label{baryon}
\end{equation}
where we use $B= \frac{8}{23}(B-L)$ which is also satisfied in this model.
The parameter $z_{\rm EW}$ is related to the sphaleron decoupling temperature 
$T_{\rm EW}$ through $z_{\rm EW}=\frac{M_3}{T_{\rm EW}}$ and
$z_{\rm EW}\simeq 20$ in the present case.  
The solution of the Boltzmann equations is given in the upper panel 
of Fig.~3 for each case given in Table 1.
The mass degeneracy $\Delta_i$ is fixed to $\Delta_i=10^{-6.5}$.
In this estimation, the $CP$ phase $\varphi_3-\varphi_{1,2}$ 
in eq.~(\ref{cp}) is chosen to make $\varepsilon$ a maximum value.  
From this figure, we find that the case (ii) gives a consistent 
solution with the required lepton number asymmetry for 
the assumed right-handed neutrino mass degeneracy.
On the other hand, in the case (i) the generated baryon
number asymmetry does not reach the required value by a small amount.
Here we note that the value of $\varepsilon$ is one order of 
magnitude larger in the case (ii) than that in the case (i).
We note that this feature is caused by the neutrino mass matrix fixed by 
$p_{1,2}$ and $q_{1,2}$.
If we use these solutions, we find that the model induces the baryon
number asymmetry in each case as
\begin{equation}
{\rm (i)}~  Y_B=2.9\times 10^{-11}, \qquad
{\rm (ii)}~  Y_B=9.7\times 10^{-11}.
\end{equation}
If the $CP$ phase takes a suitable value in the present parameter setting, 
the desirable value of $Y_B$ can be
obtained in the case (ii) even for milder degeneracy among the
right-handed neutrinos compared with the ordinary resonant 
leptogenesis \cite{resonant}.

\begin{figure}[t]
\begin{center}
\epsfxsize=8cm
\leavevmode
\epsfbox{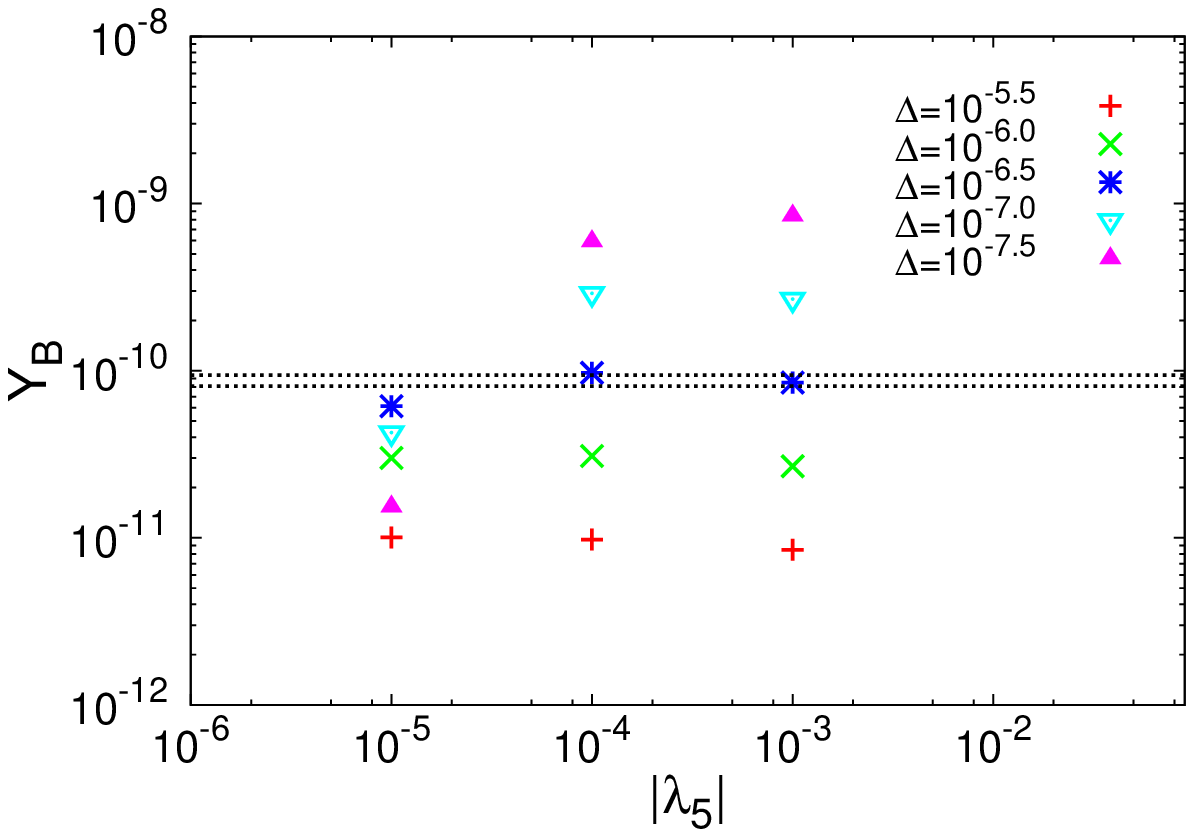} 
\end{center}
\vspace*{-3mm}
{\footnotesize {\bf Fig.~4}~~ The required mass degeneracy for the
 right-handed neutrinos to generate the sufficient baryon number 
asymmetry. We use the parameters in the case (ii) in this analysis.}
\end{figure}

It is useful to see the behavior of the lepton number violating 
reaction rate in this analysis.
The thermally averaged reaction rate is related to the reaction 
density through
\begin{equation}
\Gamma_D^{N_3}=\frac{\gamma_D^{N_3}}{n_{N_{3}}^{\rm eq}}, \qquad
\Gamma_{\rm ID}^{N_{1,2}}=\frac{\gamma_D^{N_{1,2}}}{n_\ell^{\rm eq}}, \qquad
\Gamma_N^{(2,13)}=\frac{\gamma_N^{(2,13)}}{n_\ell^{\rm eq}},
\end{equation}
for the decay of $N_{3}$ and the inverse decay of $N_{1,2}$ 
given in eq.~(\ref{decay}), and for the 2-2 scattering processes 
given in eqs.~(\ref{lv1}) and (\ref{lv2}), respectively. 
The ratio of the thermally averaged reaction rate to the Hubble 
parameter $\frac{\Gamma}{H}$ is plotted as a function of $z$ in the lower
panels of Fig.~3. These panels suggest that the lepton
number violating processes show the almost same behavior in both cases. 
We find from these panels that the lepton number violating processes 
decouple at $z~{^>_\sim}~20$ and then the washout of the generated lepton
number asymmetry is suppressed sufficiently after this period in both case.
The main reason to cause the difference in the generated lepton 
number asymmetry in both cases is considered to come from the difference
in the value of the $CP$ asymmetry $\varepsilon$, which shows the
difference of one order of magnitude between these cases as addressed already. 

We note that both the neutrino mass eigenvalues and the PMNS matrix 
does not change in this model as long as the value of $|h_{1,2}^2\lambda_5|$ 
is kept to a suitable value. If we use this feature, 
we can vary the magnitude of the
neutrino Yukawa couplings consistently with all the neutrino oscillation
data by changing the value of $|\lambda_5|$.
In Fig.~4, we show the required mass degeneracy 
$\Delta_2$ between 
the right-handed neutrinos to generate the sufficient baryon number
asymmetry for several $|\lambda_5|$ values in the case (ii).
The figure shows that the sufficient baryon number asymmetry could be
generated for $\Delta_2~{^<_\sim}~O(10^{-6.5})$ as long as $|\lambda_5|$ is 
in the region fixed by eq.~(\ref{direct1}).
Although larger $|\lambda_5|$ makes $|h_2|$ smaller and then results 
in the smaller $CP$ asymmetry, the washout due to the 
lepton number violating processes also become smaller.
In much larger $|\lambda_5|$ region, the former suppression exceeds 
the latter effect. 
On the other hand, the smaller $|\lambda_5|$ makes $|h_2|$ larger
to enhance both the $CP$ asymmetry and the washout of the generated
lepton number asymmetry.
In much smaller $|\lambda_5|$ region, the latter enhancement exceeds the
former effect. 
This explains the behavior of $Y_B$ shown in this figure. 
The sufficient baryon number asymmetry could be generated only for 
a restricted region of $|\lambda_5|$. 
It is also clear from eq.~(\ref{cp}) that the smaller
$Y_B$ is obtained for the smaller $\Delta_2$ 
at the region of $|\lambda_5|<10^{-5}$ where $\varepsilon$ is expected
to be proportional to $\Delta_2/\tilde\Gamma_{N_2}$. 
The similar behavior is obtained also in the case (i), although the
severer mass degeneracy $\Delta_1$ is required to generate the
sufficient baryon number asymmetry.

The required mass degeneracy $\Delta_i$ is much milder compared with
the resonant leptogenesis in the ordinary seesaw scenario at the TeV scale.
It is useful to remind the reader that $\Delta=O(10^{-10\sim -8})$ is 
required to generate the sufficient baryon number asymmetry there. 
This interesting feature which is also found in the normal hierarchy
case is brought by the neutrino mass 
generation mechanism in the model. 
Although the model is the simple extension of the SM,
it can consistently explain three crucial problems in the SM, that is, 
the small masses and large mixing of neutrinos, the DM
abundance and the baryon number asymmetry in the Universe, 
by fixing the parameters suitably.

\subsection{DM detection and neutrinoless double $\beta$ decay} 
Finally, we discuss the relation between the model and the low 
energy phenomena, in particular, the DM scattering with a nucleus and 
the neutrinoless double $\beta$ decay. They are considered as the
promising targets in the next generation experiments.
One might consider that eqs.~(\ref{direct1}) and (\ref{direct2}) suggest 
that the inert doublet DM might be detected in the direct search
experiments.
If $|\lambda_5|$ is small enough within the bound (\ref{direct1}),
one might expect that it is detected through the inelastic scattering. 
However, since the DM velocity has the upper bound $v_{\rm esc}$ 
as discussed before, the inelastic scattering of the inert doublet DM 
is kinematically allowed only in the restricted parameter region. 
Xenon100 has already excluded this possibility as shown in Fig.~1.

In the elastic scattering case, on the other hand, if $\lambda_3$ 
and $\lambda_4$ take suitable values within (\ref{direct2}), 
the inert doublet DM is expected to be found through the Xenon1T experiment. 
In Fig.~5, we plot the contours of
the DM-nucleon scattering cross section $\sigma_n^0$ and also the
contour of the expected sensitivity bound of Xenon1T \cite{xenon1t} in the 
$(m_\eta, \lambda_4)$ plane for both sign of $\lambda_3$.
In the left panel, we plot them for $\lambda_3=-0.4$ 
(red thin solid lines) and $-0.01$ (green thin dotted lines).  
The contour of $\sigma_n^0=4.4\times 10^{-45}$~cm$^2$ is also plotted by 
using the same type of lines as these bounds for each $\lambda_3$.
The upper shaded region could not be reached by Xenon1T and the lower
shaded regions have been excluded by Xenon100. 
We find that the Xenon100 result excludes a substantial region 
in this plane for larger value of $|\lambda_3|$.  
In the right panel, $\lambda_3$ is fixed to $1.0$ 
(red thin solid lines) and 0.4 (green thin dotted lines) and
the contours of $\sigma_n^0=2\times 10^{-45}$ and 
$4\times 10^{-46}$~cm$^2$ are plotted for each 
$\lambda_3$ in the same way as the left
panel. The shaded region cannot be reached by Xenon1T. All region in the
plane has not been excluded by Xenon100 in this case. 
In both panels, we also plot the contour of the required 
DM relic density $\Omega h^2=0.11$ by the thick lines of the same type 
as the ones used to plot the cross section for each value of $\lambda_3$.
Thus, the cross section predicted by this model should be read off 
on these lines. This figure shows that our DM candidate is expected to 
be found in the Xenon1T experiment as long as $\lambda_{3,4}$ and $m_\eta$ 
take suitable values. 
We should note that such parameters can be consistent with 
the ones which have been discussed in the previous part in relation to 
the neutrino oscillation data and leptogenesis.

\begin{figure}[t]
\begin{center}
\epsfxsize=7cm
\leavevmode
\epsfbox{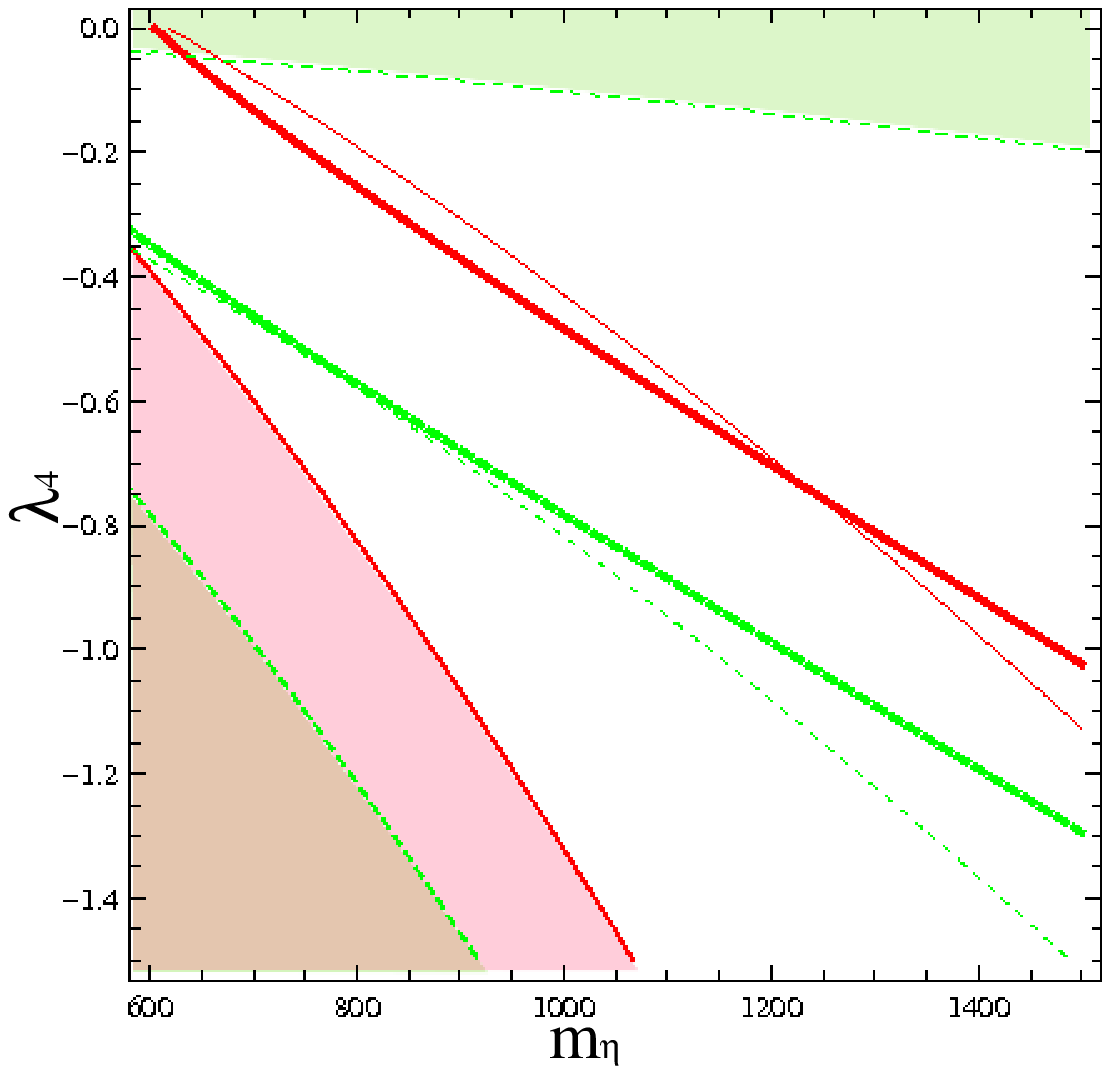} 
\hspace{5mm}
\epsfxsize=7cm
\leavevmode
\epsfbox{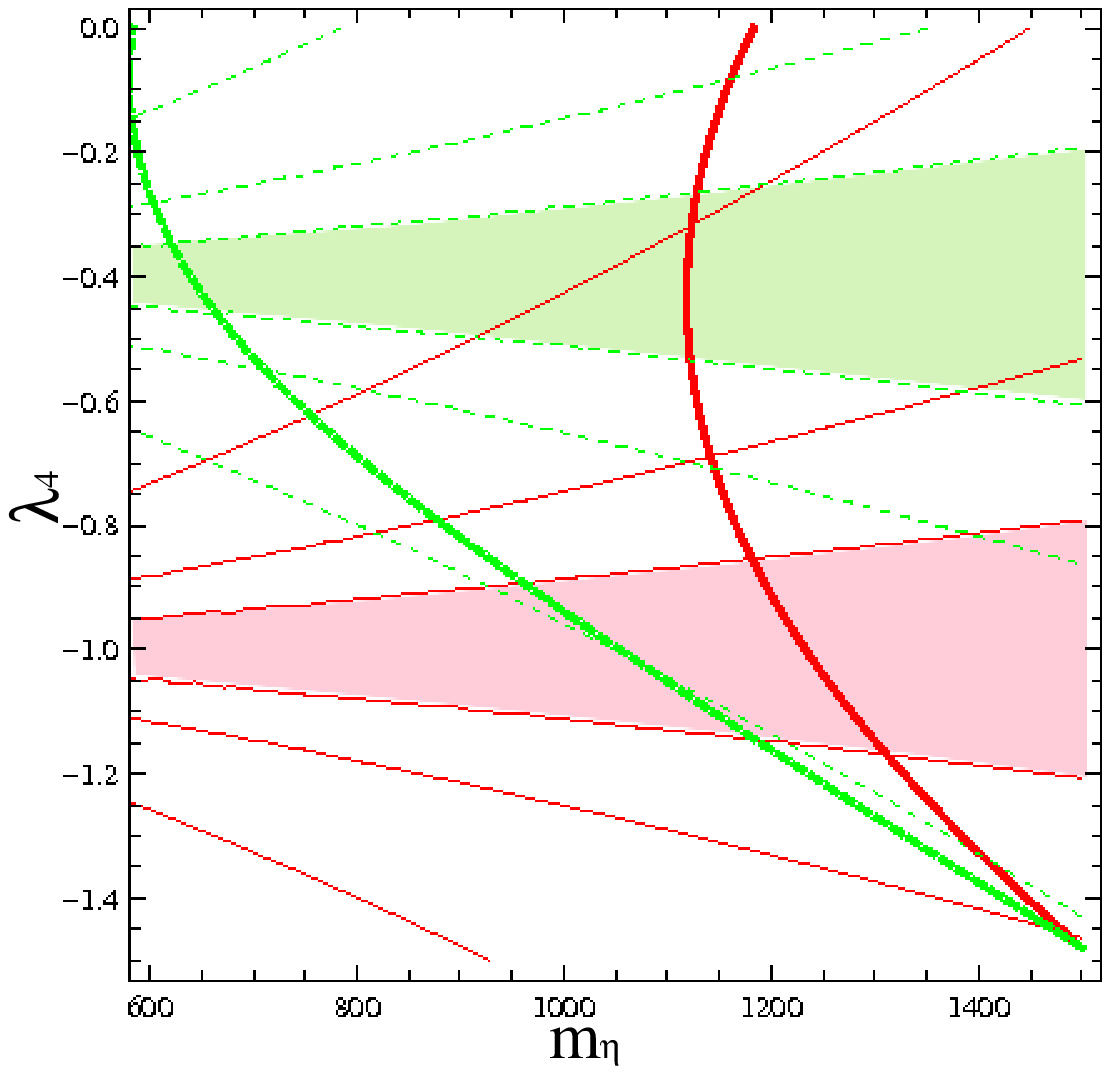} 
\end{center}
\vspace*{-3mm}
{\footnotesize {\bf Fig.~5}~~Regions in the $(m_\eta, \lambda_4)$ plane
where the DM is expected through the direct search in Xenon1T experiment.  
The sign of $\lambda_3$ is fixed to $\lambda_3<0$ in the left panel and
$\lambda_3>0$ in the right panel. The contours of $\sigma_n^0$, 
the sensitivity bound of Xenon1T, and also the contour of the DM 
relic density $\Omega h^2=0.11$ are also plotted. 
Detailed explanation of the lines can be found in the text.
In this plot, we use $|\lambda_5|=10^{-4}$.}
\end{figure}

The neutrinoless $\beta$ decay could also be another future target of
this model.  The effective mass $m_{ee}$ for 
the neutrinoless double $\beta$ decay in this model is given by
\begin{eqnarray}
m_{ee}&=&\left|\sum_{i=1}^3U_{ei}^2m_i\right|\simeq
\left|(U_{e1}^2+U_{e2}^2)\sqrt{\Delta m_{\rm atm}^2}
-U_{e1}^2\sqrt{\Delta m_{\rm sol}^2}\right| \nonumber \\
&\simeq&\sqrt{\Delta m_{\rm atm}^2}\left[ U_{e1}^4+U_{e2}^4
+2U_{e1}^2U_{e2}^2\cos (\varphi_1-\varphi_2)\right]^{1/2}.
\label{beta}
\end{eqnarray}
If we use the parameters obtained through the
previous analysis, we can estimate the value of $m_{ee}$, which is 
shown in the last column of Table 1. 
These values are obtained by changing the phases in the possible range. 
As usually expected in the inverted hierarchy case, 
we find that these values are contained in the region which could 
be observed in the next generation experiments.

Since the region of $|\lambda_5|$ is confined to rather restricted region 
through the DM search, we might have useful information on the order 
of degeneracy among the right-handed neutrinos by requiring the 
production of the observed baryon number asymmetry as
discussed above. This might allow us more detailed study of
the model by combining analysis of the neutrino oscillation data 
and the baryon number asymmetry.
On the other hand, although the effective mass for the neutrinoless 
double $\beta$ decay is related to $\lambda_5$ 
through the combination $|\lambda_5h_{1,2}^2|$ which controls the
neutrino mass eigenvalues, it is not directly related 
to $|\lambda_5|$ itself as found from eq.~(\ref{beta}).
The value of $m_{ee}$ does not change even if we change the 
value of $|\lambda_5|$ as long as the flavor structure (\ref{myukawa}) 
is kept. This is quite different from the lepton flavor violating 
processes discussed in the previous part.  
As a result, even if we use the $|\lambda_5|$ value restricted by the DM
direct search, the effective mass $m_{ee}$ is not expected to
be determined in more precise way.\footnote{We should also note 
that the $CP$ phases appeared in eqs.(\ref{beta}) and (\ref{cp}) 
have no direct relation.}
The direct DM search seems to be a promising experiment
to examine the model other than the search of the charged 
scalars in the LHC.

\section{Conclusion}
We have constructed a concrete example to realize the inverted 
hierarchy of the neutrino masses in the radiative neutrino mass model at TeV
scales. The model is an interesting and simple extension of the SM by adding an
additional doublet scalar and three right-handed neutrinos only.
In this example, we have examined the possibility for the simultaneous 
explanation of the neutrino oscillation data, the DM abundance and 
the baryon number asymmetry in the Universe.
The results show that their simultaneous explanation is possible 
as long as the DM is identified with the lightest neutral component 
of the inert doublet scalar and the resonant leptogenesis is assumed. 

In the resonant leptogenesis, the degeneracy required for the
right-handed neutrinos is milder by a few order of magnitude 
than the one known in the ordinary resonant leptogenesis at TeV scales.  
This feature is brought about as a result of the present 
neutrino mass generation
scheme. The right-handed neutrino spectrum also affects the resonant
leptogenesis. This is caused by the flavor structure of the neutrino
Yukawa couplings which is imposed by the neutrino oscillation data. 
The neutrinoless double $\beta$ decay can be observed 
through the next generation experiments also in this inverted hierarchy model. 
The most promising experiment to examine the model may be the DM 
direct search. It is expected to be found in the Xenon1T experiment.

What DM is in this radiative neutrino mass model is closely related to  
a key parameter $\lambda_5$ of the model, which is also connected 
with both the neutrino mass generation and the production of the baryon 
number asymmetry. Thus, if a DM direct search could find some
candidate, the model might be studied in more definite way based on it.

\section*{Acknowledgement}
This work is partially supported by a Grant-in-Aid for Scientific
Research (C) from Japan Society for Promotion of Science (No.24540263).

\newpage
\section*{Appendix}
We summarize the formulas of the reaction density
used in the Boltzmann equations \cite{cross} for the number density 
of $N_3$ and the lepton number asymmetry.\footnote{
These formulas are the same as the ones given in the Appendix of \cite{ks}
except that they are arranged so as to be applicable 
to the mass spectrum assumed in this paper. Typos and errors 
are corrected.}
In order to give the expression for the reaction density of the relevant
processes, we introduce dimensionless variables
\begin{equation}
x=\frac{s}{M_3^2}, \qquad a_j=\frac{M_j^2}{M_3^2}, \qquad 
a_\eta=\frac{M_\eta^2}{M_3^2},
\end{equation}
where $s$ is the squared center of mass energy.
The reaction density for the decay of $N_j$ can be 
expressed as
\begin{equation}
\gamma_D^{N_j}=\frac{(hh^\dagger)_{jj}}{8\pi^3}
M_3^4a_j\sqrt{a_j}\left(1-\frac{a_\eta}{a_j}\right)^2
\frac{K_1(\sqrt{a_j}z)}{z}, 
\label{decay}
\end{equation} 
where $K_1(z)$ is the modified Bessel function of the second kind.

The reaction density for the scattering process is expressed as
\begin{equation}
\gamma(ab\rightarrow ij)=\frac{T}{64\pi^4}\int^\infty_{s_{\rm min}}ds~
\hat\sigma(s)\sqrt{s}K_1\left(\frac{\sqrt{s}}{T}\right),
\end{equation}
where
$s_{\rm min}={\rm max}[(m_a+m_b)^2,(m_i+m_j)^2]$ and 
$\hat\sigma(s)$ is the reduced cross section. 
In order to give the concrete expression for the reaction density 
relevant to eq.~(\ref{bqn}), we define the following 
quantities for convenience:
\begin{eqnarray}
 && \frac{1}{D_i(x)}=\frac{x-a_i}{(x-a_i)^2+a_i^2c_i}, \qquad 
c_i=\frac{1}{64\pi^2}\left(\sum_{\alpha=e,\mu,\tau}
|h_{\alpha i}|^2\right)^2\left(1-\frac{a_\eta}{a_i}\right)^4, \nonumber \\
&&\lambda_{ij}=\left[x-(\sqrt{a_i}+\sqrt{a_j})^2\right]
\left[x-(\sqrt{a_i}-\sqrt{a_j})^2\right],
 \nonumber \\
&&L_{ij}=\ln\left[\frac{x-a_i-a_j+ 2a_\eta +\sqrt{\lambda_{ij}}}
{x-a_i-a_j +2 a_\eta -\sqrt{\lambda_{ij}}}\right], \nonumber \\
&&L_{ij}^\prime=\ln\left[\frac{\sqrt{x}(x-a_i-a_j-2a_\eta)
+\sqrt{\lambda_{ij}(x-4a_\eta)}}
{\sqrt{x}(x-a_i-a_j-2a_\eta) 
-\sqrt{\lambda_{ij}(x-4a_\eta)}}\right].
\end{eqnarray}

As the lepton number violating scattering processes induced through the
$N_i$ exchange,
we have
\begin{eqnarray}
\hat\sigma^{(2)}_N(x)&=&\frac{1}{2\pi}\frac{(x-a_\eta)^2}{x^2}
\left[\sum_{i=1}^3(hh^\dagger)_{ii}^2\frac{a_i}{x}
\left\{\frac{x^2}{xa_i -a_\eta^2}+\frac{x}{D_i(x)}
+\frac{(x-a_\eta)^2}{2D_i(x)^2}\right.\right.\nonumber \\
&-&\left.\frac{x^2}{(x-a_\eta)^2}
\left(1+\frac{x+a_i-2a_\eta}{D_i(x)}\right)
\ln\left(\frac{x(x+a_i-2a_\eta)}{xa_i-a_\eta^2}\right)\right\}\nonumber \\
&+&\left.
\sum_{i>j}{\rm Re}[(hh^\dagger)_{ij}^2]\frac{\sqrt{a_ia_j}}{x}\left\{
\frac{x}{D_i(x)}+\frac{x}{D_j(x)}+\frac{(x-a_\eta)^2}{D_i(x)D_j(x)}
\right.\right.\nonumber \\
&+&\left.\left.\frac{x^2}{(x-a_\eta)^2}
\left(\frac{2(x+a_i-2a_\eta)}{a_j-a_i}-
\frac{x+a_i-2a_\eta}{D_j(x)}\right)\ln\frac{x(x+a_i-2a_\eta)}{xa_i-a_\eta^2}
\right.\right. \nonumber\\
&+&\left.\left.\frac{x^2}{(x-a_\eta)^2}
\left(\frac{2(x+a_j-2a_\eta)}{a_i-a_j}-
\frac{x+a_j-2a_\eta}{D_i(x)}\right)\ln\frac{x(x+a_j-2a_\eta)}{xa_j-a_\eta^2}
\right\}\right]
\label{lv1}
\end{eqnarray}
for $\ell_\alpha\eta^\dagger \rightarrow \bar\ell_\beta\eta$ and also
\begin{eqnarray}
\hat\sigma^{(13)}_N(x)&=&\frac{1}{2\pi}
\left[\sum_{i=1}^3(hh^\dagger)^2_{ii}\left\{
\frac{a_i(x^2-4xa_\eta)^{1/2}}{a_ix+(a_i-a_\eta)^2}\right.\right. \nonumber \\
&+&\left.\left.
\frac{a_i}{x+2a_i-2a_\eta}
\ln\left(\frac{x+(x^2-4xa_\eta)^{1/2}+2a_i-2a_\eta}
{x-(x^2-4xa_\eta)^{1/2}+2a_i-2a_\eta}\right)\right\}
\right.\nonumber\\
&+& \sum_{i>j}{\rm Re}[(hh^\dagger)_{ij}^2]
\frac{\sqrt{a_ia_j}}{x+a_i+a_j-2a_\eta}
\nonumber \\
&\times& \Big\{\frac{2x+3a_i+a_j-4a_\eta}{a_j-a_i}
\ln\left(\frac{x+(x^2-4xa_\eta)^{1/2}+2a_i-2a_\eta}
{x-(x^2-4xa_\eta)^{1/2}+2a_i-2a_\eta}\right) \nonumber \\
&+&\left.\left. \frac{2x+a_i+3a_j-4a_\eta}{a_i-a_j}
\ln\left(\frac{x+(x^2-4xa_\eta)^{1/2}+2a_j-2a_\eta}
{x-(x^2-4xa_\eta)^{1/2}+2a_j-2a_\eta}\right)
\right\}\right] 
\label{lv2}
\end{eqnarray}
for $\ell_\alpha\ell_\beta \rightarrow \eta\eta$.
The cross terms has no contribution if the $CP$ phases are assumed to 
satisfy $|\sin2(\varphi_{1,2}-\varphi_3)|=1$. 
We adopt this possibility in the numerical analysis, for simplicity. 
Since another type of the lepton number violating process 
$N_{i}N_{j}\rightarrow \ell_\alpha\ell_\beta$ induced 
by the $\eta$ exchange could be suppressed for a small 
$|\lambda_5|$, we can neglect them safely for the value of
$|\lambda_5|$ used in this analysis.
 
As the lepton number conserving scattering processes which contribute to
determine the number density of $N_3$, we have
\begin{eqnarray}
\hat\sigma^{(2)}_{N_iN_j}(x)&=&\frac{1}{4\pi}
\left[(hh^\dagger)_{ii}(hh^\dagger)_{jj}
\left\{\frac{\sqrt{\lambda_{ij}}}{x}\left(1+
\frac{(a_i-a_\eta)(a_j-a_\eta)}{(a_i-a_\eta)(a_j-a_\eta)+xa_\eta}\right)
\right.\right.\nonumber \\
&+&\left.\left.\frac{a_i+a_j-2a_\eta}{x}L_{ij}\right\}
- {\rm Re}[(h h^\dagger)_{ij}^2]
\frac{2\sqrt{a_ia_j}L_{ij}}{x-a_i-a_j+2a_\eta}\right]
\label{nlv1}
\end{eqnarray}
for $N_{i}N_{j} \rightarrow \ell_\alpha\bar\ell_\beta$ which are 
induced through the $\eta$ exchange and also
\begin{eqnarray}
\hat\sigma^{(3)}_{N_iN_j}(x)&=&
\frac{1}{4\pi}\frac{(x-4a_\eta)^{1/2}}{x^{1/2}}
\left[|(hh^\dagger)_{ij}|^2\left\{
\frac{\sqrt{\lambda_{ij}}}{x}
\Big(-2  \right.\right. \nonumber \\
&+&\left.\frac{a_\eta(a_i- a_j)^2}
{(a_\eta-a_i)(a_\eta-a_j)x +(a_i-a_j)^2a_\eta}\Big) 
+\frac{x^{1/2}}{(x-4a_\eta)^{1/2}}\left(1-2\frac{a_\eta}{x}\right)
L_{ij}^\prime\right\}  \nonumber\\
&-&\left. 2{\rm Re}[(h h^\dagger)_{ij}^2]
\frac{\sqrt{a_ia_j}(a_i+a_j-2a_\eta)L_{ij}^\prime}
{x(x-a_i-a_j-2a_\eta)}\right]
\label{nlv2}
\end{eqnarray}
 for $N_iN_j \rightarrow \eta\eta^\dagger$ which are 
induced through the $\ell_\alpha$ exchange.
The cross terms in these reduced cross sections can be neglected
if the same assumption is made for the $CP$ phases 
 as eqs.~(\ref{lv1}) and (\ref{lv2}). 

\newpage
\bibliographystyle{unsrt}

\end{document}